\documentclass[sigconf, nonacm]{acmart}
\usepackage{adjustbox}
\AtBeginDocument{%
  \providecommand\BibTeX{{%
    \normalfont B\kern-0.5em{\scshape i\kern-0.25em b}\kern-0.8em\TeX}}}

\makeatletter
\def\@ACM@checkaffil{
    \if@ACM@instpresent\else
    \ClassWarningNoLine{\@classname}{No institution present for an affiliation}%
    \fi
    \if@ACM@citypresent\else
    \ClassWarningNoLine{\@classname}{No city present for an affiliation}%
    \fi
    \if@ACM@countrypresent\else
        \ClassWarningNoLine{\@classname}{No country present for an affiliation}%
    \fi
}
\makeatother

\usepackage{algorithmic}
\usepackage{booktabs,siunitx}
\usepackage{multirow}
\usepackage{hyperref}
\usepackage{gensymb}
\usepackage{amsmath}
\usepackage{xcolor}
\usepackage{xspace}
\renewcommand{\paragraph}[1]{\vspace{0.12in}\noindent{\bf{#1}.}}

\newcommand{\sysname}{{\scshape Graphene}\xspace}
\newenvironment{packeditemize}{
\begin{list}{$\bullet$}{
\setlength{\itemsep}{1.5pt}
\setlength{\labelwidth}{8pt}
\setlength{\leftmargin}{10pt}
\setlength{\labelsep}{3pt}
\setlength{\listparindent}{\parindent}
\setlength{\parsep}{1.5pt}
\setlength{\parskip}{1.5pt}
\setlength{\topsep}{1.5pt}}}{\end{list}}

\begin{document}


\title{\sysname: Infrastructure Security Posture Analysis with AI-generated Attack Graphs}





\author{Xin Jin}
\authornote{These authors contributed equally to this work.}
\authornote{This work was partially completed during the authors' internship at Cisco Research.}
\affiliation{%
	\institution{The Ohio State University}
}
\email{jin.967@osu.edu}

\author{Charalampos Katsis}
\authornotemark[1]
\authornotemark[2]
\affiliation{%
	\institution{Purdue University}
}
\email{ckatsis@purdue.edu}

\author{Fan Sang}
\authornotemark[1]
\authornotemark[2]
\affiliation{%
	\institution{Georgia Institute of Technology}
}
\email{fsang@gatech.edu}

\author{Jiahao Sun}
\authornotemark[2]
\affiliation{%
	\institution{Georgia Institute of Technology}
}
\email{jiahaosun@gatech.edu}

\author{Elisa Bertino}
\affiliation{%
	\institution{Purdue University}
}
\email{bertino@purdue.edu}

\author{Ramana Rao Kompella}
\affiliation{%
	\institution{Cisco Research}
}
\email{rkompell@cisco.com}

\author{Ashish Kundu}
\affiliation{%
	\institution{Cisco Research}
}
\email{ashkundu@cisco.com}

\begin{abstract}
The rampant occurrence of cybersecurity breaches imposes substantial limitations on the progress of network infrastructures, leading to compromised data, financial losses, potential harm to individuals, and disruptions in essential services. The current security landscape demands the urgent development of a holistic security assessment solution that encompasses vulnerability analysis and investigates the potential exploitation of these vulnerabilities as attack paths. In this paper, we propose \textit{\sysname}, an advanced system designed to provide a detailed analysis of the security posture of computing infrastructures. Using user-provided information, such as device details and software versions, \sysname performs a comprehensive security assessment. This assessment includes identifying associated vulnerabilities and constructing potential attack graphs that adversaries can exploit. Furthermore, \sysname evaluates the exploitability of these attack paths and quantifies the overall security posture through a scoring mechanism. The system takes a holistic approach by analyzing security layers encompassing hardware, system, network, and cryptography. Furthermore, \sysname delves into the interconnections between these layers, exploring how vulnerabilities in one layer can be leveraged to exploit vulnerabilities in others. In this paper, we present the end-to-end pipeline implemented in \sysname, showcasing the systematic approach adopted for conducting this thorough security analysis.
\end{abstract}

\maketitle

\section{Introduction}

The escalating complexity of enterprise networks has resulted in environments highly susceptible to cyber-attacks~\cite{cybercrime2022}. The proliferation of thousands of applications deployed on network hosts and the influx of software packages from diverse sources, whether active or dormant, significantly magnify security concerns.  
The presence of such a diverse range of software raises questions about the potential vulnerabilities they may introduce to the network. 
The intricate network of interconnected devices, coupled with emerging technologies such as cloud computing, Internet of Things (IoT) devices, and interconnected systems, expands the attack surface exponentially, providing attackers with numerous entry points~\cite{jin2022edge}. Moreover, the interconnected nature of networks intensifies the impact of a single vulnerability, enabling adversaries to navigate interconnected systems and compromise multiple hosts and devices. Finally, hundreds of vulnerabilities are disclosed monthly in the national vulnerability databases; thus, an approach to evaluating and understanding their impact is of paramount importance for several reasons, such as prioritizing patching efforts. 

Faced with those complex challenges, ensuring a robust security posture is thus critical in today's dynamic landscape of enterprise networks, where strategies, policies, and practices collectively fortify an organization's cybersecurity defense.

\noindent \textbf{Problem Scope.} Given the expanding attack surface, 
we need comprehensive systems able not only to identify vulnerabilities specific to the infrastructure of interest but also to discern how these vulnerabilities can be exploited in sequences. Our goal is to develop a solution that leverages \textit{attack graphs} to (1) understand how vulnerabilities might serve as a sequence of steps in a multi-step attack and (2) scrutinize and grasp the implications of a vulnerability on the infrastructure under analysis. The system implementing the solution should incorporate continuous monitoring to detect known vulnerabilities across host applications, devices, and their configurations. Additionally, a deep understanding of the nature and impact of each vulnerability is essential to tailor effective approaches for mitigation by subject matter experts. 

\noindent \textbf{Challenges.} Designing such a system requires addressing the following challenges: \textbf{(C1)} The vulnerabilities are typically described in natural language (i.e., common vulnerabilities and exposures -- CVE) rather than in a formally defined encoded format. Therefore, a systematic approach is needed to capture the vulnerability semantics and convert them into a suitable format for further analysis. \textbf{(C2)} The formulation of attack paths constituting a chain of vulnerabilities to the attacker's objectives with the least manual effort requires extracting the conditions that allow one to exploit a vulnerability (i.e., \textit{preconditions}) and the state of the system once a vulnerability is exploited (i.e., \textit{postconditions}). \textbf{(C3)} Finally, it is hard to come up with security quantification metrics that capture both the criticality of vulnerabilities and the impact on the system under analysis.

\noindent \textbf{Our Approach.} To address such challenges, we design an innovative solution, called \textit{\sysname}, which serves as a comprehensive security posture analyzer for computing infrastructures and applications. 
\sysname operates seamlessly by continuously monitoring trustworthy data sources, such as national vulnerability databases, for disclosed vulnerabilities specific to the devices and configuration of a given infrastructure. 

Using named entity recognition (NER), \sysname is able to extract the semantic meaning of these vulnerabilities and encode them into a latent space for in-depth analysis. In particular, \sysname automatically extracts the preconditions required for an adversary to exploit a vulnerability and the result after exploiting the vulnerability (i.e., postconditions). For example, \sysname can discern that exploiting a particular CVE might require the presence of the TensorFlow XLA compiler in the Google TensorFlow version prior to 1.7.0 as a precondition. Simultaneously, \sysname captures postconditions such as a system crash resulting from a heap buffer overflow or reading from other parts of the process memory. 

Subsequently, \sysname employs a semantic similarity-matching approach based on word embeddings~\cite{li2018word} to evaluate whether the postconditions of a vulnerability match the preconditions of another, thus constructing potential attack graphs for a given topology. Once generated, \sysname analyzes these attack graphs to evaluate the security posture. We propose a novel layered approach to this analysis, where vulnerabilities within each layer share similar characteristics and can be comprehensively assessed by subject matter experts. \sysname categorizes vulnerabilities in the generated attack graphs into distinct layers, such as machine learning, system, hardware, network, and cryptography, using keyword matching and common weakness enumeration (CWE) information within the vulnerability disclosure. Hence, \sysname performs similarity matching among vulnerabilities within the same layer and across different layers, revealing how vulnerabilities in one layer can lead to the exploitation of vulnerabilities in different layers. 
This approach allows for prioritized risk analysis, mitigation strategies, and patching efforts based on the specific nature and severity of vulnerabilities at each layer. 

As a result, \sysname automatically produces two types of attack graphs: \textit{cumulative} (or multi-layer) attack graphs and  \textit{layered} attack graphs. Cumulative attack graphs show how an attacker could exploit Common Vulnerabilities and Exposures (CVEs) across multiple layers. In contrast, layered attack graphs focus on the exploitation of vulnerabilities within the same layer. This dual representation provides a comprehensive understanding of the potential attack paths across different layers and within individual layers.

By traversing the generated graphs, \sysname thoroughly analyzes the security posture of the entire infrastructure, providing both cumulative and layer-wise security analytics. These analytics reflect (1) the exploitability efforts required by an adversary for a step or a sequence of steps and (2) the impact of attack steps based on the criticality of the resources affected in the analyzed infrastructure. For instance, resources considered critical in a specific deployment substantially increase the impact of an attack. Once scores are computed with respect to such criticality, \sysname identifies vulnerabilities or highly exploitable and high-impact attack paths that require immediate attention to safeguard the system from potential attacks. 

We make the following contributions:
\begin{itemize}
    \item We introduce a novel fully-automated security posture analyzer designed to generate attack graphs for computing infrastructures.
    \item We propose a natural language processing approach based on NER and word embeddings that streamlines pre- and postcondition extraction, facilitating the generation of attack graphs without the need for manual intervention.
    \item Our framework adopts a comprehensive strategy for analyzing security postures in a multi-layered fashion. The approach analyzes each layer separately and combines them into one unified analysis.
    \item We propose risk scoring methods for tailored analysis of the underlying network infrastructure.
\end{itemize}

\paragraph{Roadmap}
Section~\ref{sec:motivation} motivates \sysname.
Section~\ref{sec:background} provides background information on vulnerability disclosures and on natural language processing using NER. 
Section~\ref{sec:sys_overview} gives an overview of the \sysname pipeline, while in Section~\ref{sec:attck_graph_generation}, we delve into the technical details, starting from the ML-based processing to the attack graph generation. Subsequently, Section~\ref{sec:risk} outlines the risk analysis methods incorporated in \sysname, designed to evaluate the security posture of a given infrastructure. In Section~\ref{sec:implementation}, we provide a comprehensive discussion of the implementation details, followed by the presentation of the evaluation results in Section~\ref{sec:eval}. Finally, Section~\ref{sec:related} discusses the related work and Section~\ref{sec:conclusion} concludes the paper.
\section{Motivation}
\label{sec:motivation}

In this section, we delve into the rationale behind employing attack graphs as a fundamental tool for conducting posture analysis. 
We then discuss limitations of previous approaches - limitations that are addressed by \sysname.

\noindent \textbf{Security Posture Analysis Using Attack Graphs.} To achieve a holistic security assessment in complex systems, attack graphs can serve as a valuable tool. 
They provide a comprehensive representation of the system's security landscape, accounting for the intricate interdependencies among its diverse components and potential pathways for attacks. 
On the one hand,  attack graphs give a holistic overview of the prevailing threats confronting the system, providing security professionals with a streamlined understanding of the overall security posture. On the other hand, attack graphs can facilitate the creation of simulation environments, enabling the exploration of hypothetical scenarios when addressing threats. Such a simulation environment essentially functions as a ``threat lab,'' empowering security professionals to optimize resource allocation for mitigating threats, thus making the threat intelligence processes more cost-effective.

Additionally, security professionals can simulate diverse scenarios by manipulating the attack graphs. For instance, the deliberate disconnection of paths through the removal of a node or edge offers insights into the potential value of investing resources to address specific threats. Conversely, the addition of a node or edge triggers the automatic inference of corresponding graph components, thereby showing the repercussions of novel threats. This proactive approach aids in better preparation for future challenges and the formulation of response defensive actions and protocols. In instances where security professionals possess insights beyond publicly available information, they can update the details of a node or edge, infusing human experience into the threat analysis process and fostering a more precise representation of the actual system.

\noindent \textbf{Limitations of Prior Works.} While prior research has explored various approaches to generate attack graphs, their practical application is limited by several factors. Some approaches require providing the relevant infrastructure-related Common Vulnerabilities and Exposures (CVEs), which in turn may require manual input~\cite{aksu2018automated, bezawada2019agbuilder, ibrahim2019attack}. However, acquiring information on infrastructure vulnerabilities must be an ongoing process, as new vulnerabilities are continually disclosed in public databases. A single vulnerability can substantially alter the security landscape by introducing new attack possibilities or high-risk pathways.

Automation is another significant limitation. Some approaches mandate that vulnerabilities be defined using proprietary formal expressions~\cite{fang2022iota, aksu2018automated, bezawada2019agbuilder}. These formal definitions may be highly specific to a particular domain and must be regularly updated to accommodate emerging attack vectors. The quality of the resulting attack graph is intrinsically linked to the accuracy of these definitions. \looseness=-1

Moreover, a critical issue in attack graph generation is the reliance on hard-coded heuristics and keyword matching to identify pre- and postconditions, which are essential for connecting CVEs within the attack graph~\cite{aksu2018automated, fang2022iota, urbanska2012structuring, ibrahim2019attack}. Those approaches may not be widely applicable to all CVEs and require continuous updates. Additionally, they cannot capture semantic information embedded in natural language text, which is essential for reasoning about the interconnection of CVEs.

Finally, many of the suggested risk analysis methods neglect to assess risk with respect to the underlying infrastructure~\cite{lu2009ranking, idika2010extending,liu2005network,abraham2014cyber}. 
This represents a notable limitation, given that the consequences of an adversary exploiting a vulnerability in a critical resource, such as a database containing sensitive client information, are considerably more severe than those associated with a less significant resource.
Consequently, conducting a vulnerability assessment tailored to the specifics of the infrastructure is imperative for fully understanding and addressing its potential impact.

\section{Background}
\label{sec:background}

\subsection{CVE and CWE}

Common Vulnerabilities and Exposures (CVE) disclosures form a critical aspect of the cybersecurity landscape, offering a standardized approach to identifying and cataloging known vulnerabilities. Assigned unique identifiers, CVEs enable effective communication and collaboration within the cybersecurity community, facilitating the timely detection and mitigation of potential threats. Concurrently, the Common Weakness Enumeration (CWE) system complements CVE by classifying and categorizing the underlying weaknesses that lead to vulnerabilities, providing a structured framework for understanding and addressing security issues. Adding to this ecosystem is the Common Vulnerability Scoring System (CVSS), which assigns numerical scores (ranging from 0 to 10) to CVE entries, quantifying the severity of vulnerabilities based on factors such as impact on confidentiality, integrity, and availability.

\subsection{Attack Graphs}
\label{sec:notions}


\begin{figure}[]
\centering
\includegraphics[width=0.45\textwidth]{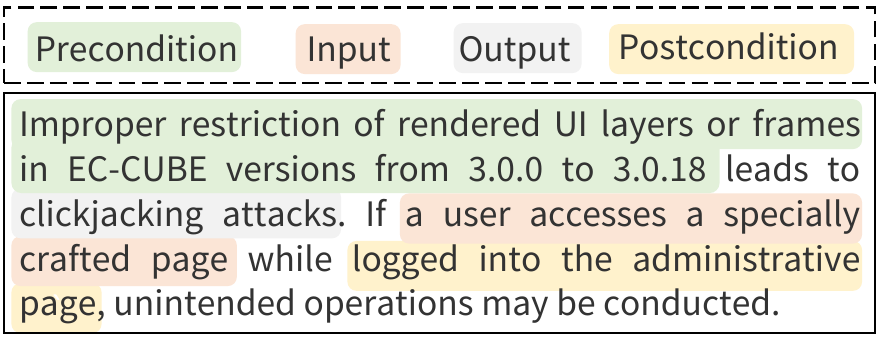}
\caption{An example of Vulnerability Description and Attack Graph Node Attributes for CVE-2020-5679.}
\label{fig:ner-example}
\end{figure}

At the high level, an attack graph can be defined 
as a structured representation of the potential paths an attacker can take to compromise a network or system by exploiting vulnerabilities~\cite{idika2010extending,aksu2018automated}. In such a representation, 
the \textit{attack graph nodes} are the attack units that
are the assembly of basic vulnerability attributes combined as coherent entities.
These attributes encompass preconditions, postconditions, inputs, and outputs. We give the notion of each attribute as follows.

\begin{packeditemize}
    \item \textbf{Precondition}. Preconditions refer to a collection of system properties that must hold for an exploit to succeed. 
    If these preconditions are met, it becomes possible to render all subsequent steps of an attack ineffective.
    \item \textbf{Postcondition}. 
    Postconditions are the system properties that hold as the results of an attack step, which are necessary for the generation of outputs.
    \item \textbf{Input}. The inputs are the actions that attackers need to take to trigger the vulnerability and perform the exploit.
    For instance, a victim software declares a fixed-length buffer to exploit a buffer overflow vulnerability.
    The input for such a vulnerability is access to the buffer elements that exceed the buffer length.
    The vulnerable system with the attack precondition can still execute normally without attack inputs. 
    For example,  the victim program with a boundary check to prevent buffer overflow is still safe if the adversary does not perform the out-boundary element operations.
    \item \textbf{Output}. 
    The outputs refer to the final values or results that the system returns or produces when exploits to vulnerabilities are executed. 
    For example, the output for the buffer overflow vulnerability can be a system crash, abnormal privilege escalation, remote code execution, etc.
\end{packeditemize}

Attack graph nodes can be categorized as attackers, attack targets, and vulnerabilities. 
Generally, attackers act as the source or root nodes, vulnerabilities function as intermediate nodes, and attack targets represent the sink/leaf nodes. 

By exploiting vulnerabilities, attackers can perform a sequence of steps or actions to attack the victims, e.g., gaining unauthorized access to a network or system.
Such steps and actions are the edges in the attack graph, serving as the basic connections.
Formally, \textit{attack graph edges} represent the transitions and chains of vulnerabilities~\cite{payne2019secure}. 
Successful attacks often require the execution of a series of exploits in a specific sequence.
For example, 
in order to compromise the macOS Kernel through Safari, the exploit of a chain of six vulnerabilities is required~\cite{jin:pwn2own2020-safari}.
Therefore, the edge connecting two nodes in an attack graph indicates that the vulnerability exploited by one node can serve as the input and trigger for the other node.

\subsection{Named Entity Recognition}
Named entity recognition (NER) involves the identification of various segments of information within a text and subsequently categorizing them into predefined classes~\cite{nadeau2007survey}. 
These classes can encompass a wide range of entities, including but not limited to individuals, organizations, and geographical regions~\cite{shishtla2008experiments}. 
In its early stages, NER heavily relied on deterministic rules~\cite{grishman1996message}. 
However, this rule-based approach often fell short of the desired efficiency standards~\cite{li2020survey}. 
One significant breakthrough came with machine learning algorithms, which introduced probabilistic approaches to entity recognition~\cite{li2020survey,xu2019document,hu2023zero}.
For instance, popular learning-based NER techniques include the hidden Markov model~\cite{zhou2002named} and conditional random field~\cite{shishtla2008experiments}. 
More recently, deep learning methodologies have played a pivotal role in enhancing the performance of NER systems. Notably, Xu et al. introduced a novel approach by combining the bidirectional LSTM model with CRF~\cite{xu2019document}.

\subsection{Word Embeddings}
Word embedding is a natural language processing technique that represents words in a high-dimensional vector space, where each dimension corresponds to a specific feature of the word~\cite{mikolov2013distributed,mikolov2013efficient}. 
The vector values are learned through a mathematical process that analyzes the co-occurrence patterns of words in a large corpus of text data so that words that occur in similar contexts are mapped to vectors that are closer together in the vector space, while words that are dissimilar in context are mapped to vectors that are farther apart.
Many pre-trained word vectors, such as word2vec and Glove~\cite{naili2017comparative,pennington2014glove}, have been used in a variety of natural language processing tasks, such as text classification, machine translation, and named entity recognition. \looseness=-1

\section{System Overview}
\label{sec:sys_overview}

In this section, we present an overview of the \sysname pipeline, followed by a discussion of the system's architecture.

\begin{figure}[]
\centering
\includegraphics[width=1\columnwidth]{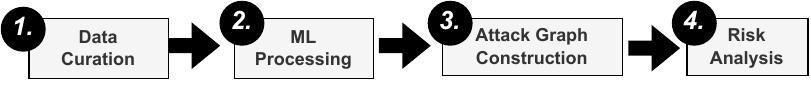}
\caption{\sysname' pipeline.}
\vspace{-0.2in}
\label{fig:pipeline}
\end{figure}

Figure~\ref{fig:pipeline} provides a comprehensive overview of the \sysname pipeline, encompassing the following key steps:

\noindent\textbf{(1) Data Curation:} This initial stage involves utilizing user-provided details about the network infrastructure, including information on network topology, communicating entities, and device specifications. The system then queries the national vulnerability database to retrieve Common Vulnerabilities and Exposures (CVE) disclosures that are specific to the installed components, such as applications and devices.

\noindent\textbf{(2) ML Processing:} In this phase, the system leverages machine learning (ML) techniques to extract contextual vulnerability information from natural language text automatically. This includes capturing pre- and postconditions. The system employs word embedding techniques to establish vulnerability similarity based on the extracted conditions. Additionally, it classifies each vulnerability node to its corresponding layer based on keyword matching and CWE information provided within the CVE disclosure.

\noindent\textbf{(3) Attack Graph Construction:} In this phase, \sysname constructs attack graphs by integrating the attack graph nodes identified in step 1 with the vulnerability similarity results obtained in step 2.

\noindent\textbf{(4) Risk Analysis:} Finally, \sysname conducts a thorough security posture analysis based on the attack graphs generated by the previous step. \sysname traverses the graphs, utilizing various vulnerability parameters such as exploitability and risk scores. This quantification process is essential for assessing the security posture layer by layer and cumulatively. Additionally, it helps identify the most impactful attack paths within the network infrastructure.

\section{Automated Attack Graph Generation} 
\label{sec:attck_graph_generation}


In this section, we present our approach to building attack graphs based on machine learning and natural language processing. 
The first step of the approach is the identification of the fundamental components of attack graphs, i.e., graph nodes,  
by using named entity recognition -- discussed in Section~\ref{sec:ner}. The second step is to establish edges between attack graph nodes by modeling preconditions, postconditions, inputs, and outputs using word embeddings -- discussed in Section~\ref{sec:embedding}.

\subsection{Attack Graph Node Identification} \label{sec:ner}

The attack graph nodes are the fundamental element of attack graphs, where they provide relevant vulnerability details, i.e., preconditions, postconditions, inputs, and outputs.
In contemporary computer systems, numerous attack surfaces exist, amplifying the scope of extensive attack graph nodes. As highlighted in Section~\ref{sec:motivation}, current approaches have significant shortcomings in scalability and adaptability.
For instance, a previously proposed approach to identify attack graph nodes~\cite{payne2019secure} relies on network sniffing to detect these nodes, which only applies to IoT networks.
We address such limitations by analyzing vulnerability descriptions and reports, thus enabling large-scale comprehensive analyses.   \looseness=-1

Identifying 
attack graph nodes from vulnerability descriptions and reports is not trivial, as it requires understanding the semantics of vulnerabilities. Consider the description of CVE-2020-5679 shown in~\autoref{fig:ner-example}, 
which can be exploited to perform clickjacking attacks.
With manual efforts, human analysts can identify the key properties of this vulnerability by understanding the description and identifying the words or phrases that represent the nodes.
For example, the precondition of CVE-2020-5679 consists of the target version of EC-CUBE with the improper restriction of UI layers.
To identify such a precondition, we need to interpret the semantics of this description.
A naive approach is to manually identify it.
However, manual identification of graph nodes based on human knowledge is not scalable.
Another approach 
is to use pivot words~\cite{urbanska2012structuring} or heuristic rules~\cite{aksu2018automated,ibrahim2019attack,payne2019secure,feng2019understanding,bridges2013automatic}.
For example, the word ``by'' is used to identify vulnerability conditions as pivot words~\cite{urbanska2012structuring}.
However, these methods are not generalizable since natural language is noisy~\cite{jin2022symlm}, e.g., different vulnerability descriptions can include semantic but syntactically different expressions.

\begin{figure}[]
\centering
\includegraphics[width=.95\columnwidth]{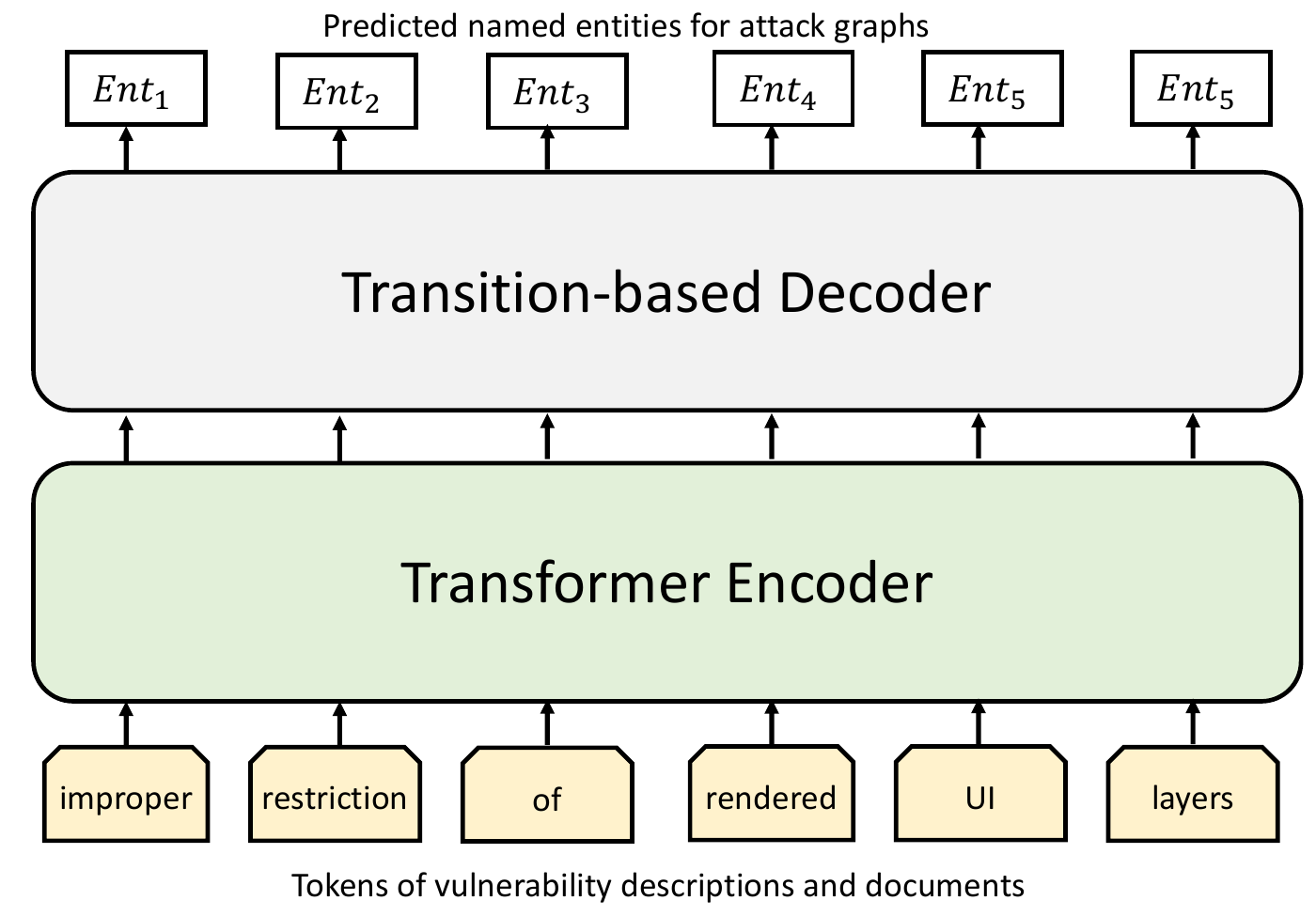}
\vspace{-0.1in}
\caption{The Named Entity Recognition Model.}
\vspace{-0.2in}
\label{fig:ner-model}
\end{figure}

\noindent\textbf{Node Identification by Entity Recognition.}
We propose to use NER for automatic attack graph node identification, which is a process of identifying entities in input texts by classifying words and phrases in vulnerability texts into the corresponding entities (addressing \textbf{C1}).
While existing approaches have applied NER to analyze security reports~\cite{weerawardhana2014automated, binyamini2021framework,jin2022understanding}, they either only focus on specific domains, e.g., home computers~\cite{binyamini2021framework}, or use weak models~\cite{jin2022understanding} (see Section~\ref{sec:related} for details).
Observing such limitations, we propose identifying security entities with a more general scope and using generative language models.
In \autoref{fig:ner-model}, our model takes as input the vulnerability descriptions and documents, which are tokenized into natural language tokens. 
The transformer encoder is a pre-trained model (e.g., roBERTa model~\cite{liu2019roberta}), which has been pre-trained on a very large corpus.
The encoder generates semantic embeddings by mapping the input tokens into latent space.
The transition-based decoder maps the token embeddings into named entities based on a finite-state transducer~\cite{kuhlmann2011dynamic}.

In addition, NER involves the other crucial task of identifying the entity type and semantics. 
However, existing solutions focus on entities like software versions and network ports~\cite{binyamini2021framework}, which are often limited to particular domains and lack general applicability.
Meanwhile, the MITRE CVE project released the official template for \textit{all CVE descriptions} that covers the critical details for automated phrasing~\cite{mitre}.
Therefore, we define six entities based on this template: vulnerability type, affected product, root cause, impact, attacker type, and attack vector, which are general to vulnerabilities because of the wide adoption of the MITRE CVE template. 
After identifying the various entities, we systematically categorize them as attributes of attack nodes, namely preconditions, postconditions, inputs, and outputs. 
In this framework, we classify the affected product entity as a precondition. The vulnerability type is categorized as a postcondition. Attacker type and root cause are considered inputs. 
Finally, the impact and attack vector entities are categorized as outputs.
In cases where certain vulnerabilities, such as public CVEs, have been assigned manual evaluation scores such as exploitability, severity, and impact score, we can also incorporate these scores as a ``golden standard'' for the corresponding attack graph nodes, if available. This additional information enhances the precision and accuracy of the attack graph by incorporating established metrics for attack graph analysis.

\subsection{Attack Graph Edge Connection} \label{sec:embedding}

The edge connection between two graph nodes indicates that the vulnerability exploited by one node can serve as the input and trigger for the other node.
To establish connections between attack nodes, it is necessary to identify pairs of nodes whose input and output ports can be matched under the same preconditions and postconditions.
Since such node attributes are identified from natural language descriptions, a simple way to build edges is to match the words of different graph nodes.
However, such a solution may not always work since morphological words (e.g., synonyms, abbreviations, and misspellings) are widely used in vulnerability descriptions and texts.
For instance, in the case of web injection vulnerabilities, the abbreviation ``XSS'' is frequently employed to refer to cross-site scripting.
To overcome such limitations, we propose to utilize word embeddings to semantically match attack graph nodes and construct attack edges to address the requirement of precise node matching.

\noindent\textbf{Word Embedding-based Edge Construction.}
In natural languages, words in different contexts can carry different meanings.
For example, the word ``band'' has different meanings under material and music context and corpus. In other words, the word embeddings trained on one specific domain cannot be directly used in other domains because there can be semantic changes.
Therefore, we argue that existing word vectors (e.g., Glove and word2vec~\cite{naili2017comparative,pennington2014glove}), which are pre-trained on general English corpus (e.g., Wikipedia), cannot precisely deliver word semantics in the security domain.

In order to obtain embeddings tailored for holistic security assessment, we propose to train embedding models using a corpus specifically focused on security (addressing \textbf{C2}). This approach offers two distinct advantages: (1) The resulting word embeddings will facilitate the accurate semantic matching of attack nodes, allowing for precise identification and classification.
(2) By quantifying the matching outcomes as similarity scores, we can assign weights to the attack edges, enabling a more nuanced representation of the severity or relevance of each attack.
To obtain the similarity scores, we first compute the semantic representations of attack node attributes (e.g., preconditions, postconditions, inputs, and outputs) by averaging the vectors of all words in the ports.
Next, we calculate the node similarity by the cosine similarity of the attribute semantic representations.

\begin{figure}[]
\centering
\includegraphics[width=.9\columnwidth]{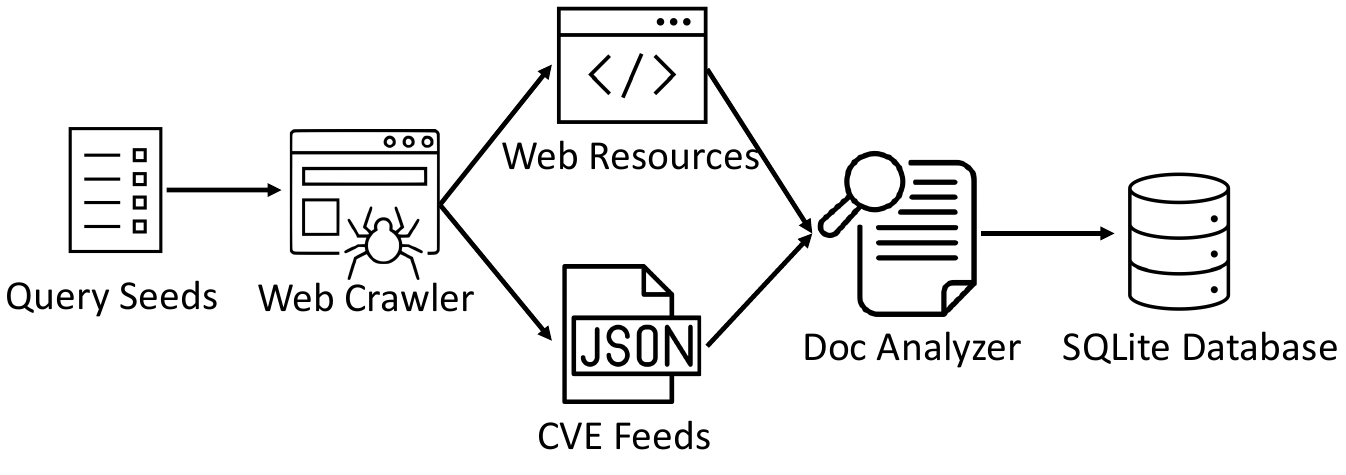}
\caption{The Security Corpus Curation Framework.}
\label{fig:crawler-framework}
\end{figure}

The key task for the word embeddings for \sysname is to curate the security corpus that is general to different vulnerabilities and attack scenarios. 
As discussed in Section~\ref{sec:background}, the NVD database contains detailed vulnerability descriptions and references of CVEs that have undergone thorough manual assessment and processing~\cite{booth2013national}.
This has resulted in the generation of succinct information regarding various vulnerabilities.
For the CVE description collection, we download CVE feeds hosted on the NVD website, which package CVE entries with descriptions.
Moreover, there are also many other online resources, such as MITRE ATT\&CK knowledge base~\cite{strom2018mitre}, that collect comprehensive vulnerability information such as Common Vulnerability and Exposures (CWEs), which we also consider in our dataset curation process.

We build a security corpus curation framework as shown in \autoref{fig:crawler-framework}.
The framework takes as input a list of query seeds (e.g., CVEs and keywords).
The output of the web crawler includes online web pages and resources and the CVE feeds.
Note that the CVE feeds contain much useful information, such as CWEs, vulnerable products, affected versions, CVSS scores, and reference links~\cite{fan2020ac}.
After obtaining all the online resources and CVE feeds, we run the doc parser to clean up these documents and extract useful content, such as the CVE descriptions and vulnerability texts.

\begin{figure}[]
\centering
\includegraphics[width=.9\columnwidth]{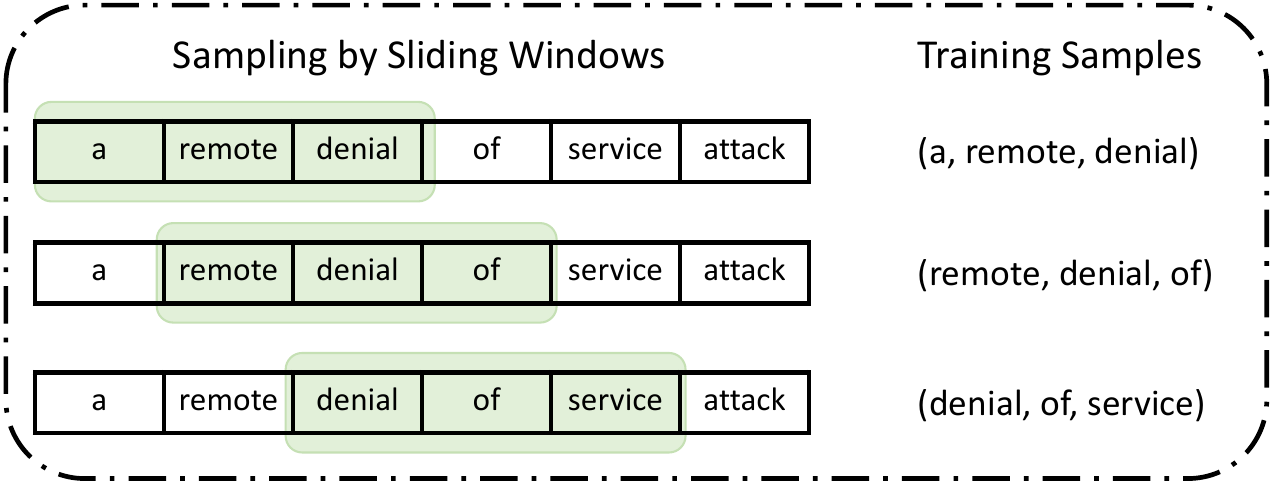}
\caption{The Dataset Sampling and Training Samples.}
\label{fig:dataset-sample}
\end{figure}

To clean our curated dataset, we first process the dataset by removing punctuation.
We do not follow the common natural language processing practice of removing stop words, as some of these words play a key role in the security context.
For example, removing the stop word ``of'' from the phrase ``denial of service'' will break this popular attack phrase and change its semantics.
Next, we sample the processed security documents into training samples using sliding windows as shown in \autoref{fig:dataset-sample}.
The training samples are sets of center words and context words.
For example, in the word set \texttt{(a, remote, denial)}, \texttt{remove} is the center word while \texttt{a} and \texttt{denial} are the context words.
This sampling method enables us to learn the word semantics by modeling the context information.
For example, the word embedding model Continuous Bag-of-Words (CBOW)~\cite{mikolov2013efficient} learns word semantics by optimizing the probability likelihood estimation:

\begin{equation}
    \frac{1}{N} \sum_{i=1}^{N}\sum_{-c\leq j\leq c, j\neq 0} log(p(w_i\mid w_{i+j}))
\end{equation}

where $N$ is the total number of words in the security document, $c$ is the sliding window size, and $w_i$ and $w_{i+j}$ are the center word and context word.

\subsection{Attack Graph Construction and Partition}

The attack graph comprises three primary node types: the attacker (source nodes), existing CVEs associated with different network entities (intermediate nodes), and CWEs serving as attacking targets (sink nodes).
The construction of edges connecting nodes and the assignment of corresponding edge weights depend on the types of connected nodes and available data. By default, an edge is established from the attacker node to each CVE node, assuming the attacker can exploit the CVE. This inclusivity aligns with \sysname' objective to generate all conceivable scenarios under various adversarial assumptions in the network. For example, if a CVE's attack vector necessitates physical access to the device, \sysname includes an analysis of such a scenario. Users can subsequently filter the generated attack graphs based on their network's specific adversarial assumptions. The edge weights for these edges are predominantly determined by CVSS base scores, indicating the likelihood of a successful attacker exploit, directly correlating with the severity of the threat posed by exposure to such vulnerabilities.

For any given pair of CVE nodes, an edge is established if the postcondition of one CVE node aligns (partially) with the precondition of another CVE node, indicating the potential for an attacker to exploit one vulnerability to access the other—essentially forming a chain of vulnerabilities for exploitation. Beyond relying on CVSS scores, the weights assigned to these edges are contingent on the node matching score of the two nodes, derived from the word-embedding-based node matching. The node matching score gauges the semantic similarity between the postcondition and precondition of the directed edge connecting two CVE nodes, offering insights into the difficulty an attacker might face in reaching the latter CVE by exploiting the former. Additionally, to mitigate graph complexity arising from marginally related CVE nodes, the user can stipulate the construction of edges only when the node matching score exceeds a specified threshold. Consequently, the resulting attack graph comprises paths that are more feasible for attackers, optimizing computational resources by excluding less viable paths. 
Lastly, each CVE node is linked to CWE nodes, representing the system's encountered threats. In the CVSS database, each CVE is associated with one or several corresponding CWEs, each instantiated as an edge in the attack graph. Similar to other edge types, the CVSS scores determine the edge weights, with the option to prune edges by setting a weight threshold.

\begin{table}[]
\centering
\caption{The "Keywords" column corresponds to high-frequency keywords found in network vulnerability listings. The "Protocol" column corresponds to communication protocols between entities.}
\begin{adjustbox}{width=0.6\columnwidth,center}
\begin{tabular}{|l|ll}
\cline{1-1} \cline{3-3}
\multicolumn{1}{|c|}{\textbf{Keywords}} & \multicolumn{1}{l|}{} & \multicolumn{1}{c|}{\textbf{Protocols}} \\ \cline{1-1} \cline{3-3} 
access control         & \multicolumn{1}{l|}{} & \multicolumn{1}{l|}{tls}   \\ \cline{1-1} \cline{3-3} 
authentication         & \multicolumn{1}{l|}{} & \multicolumn{1}{l|}{ssl}   \\ \cline{1-1} \cline{3-3} 
authenticity           & \multicolumn{1}{l|}{} & \multicolumn{1}{l|}{tcp}   \\ \cline{1-1} \cline{3-3} 
authorization          & \multicolumn{1}{l|}{} & \multicolumn{1}{l|}{ip}    \\ \cline{1-1} \cline{3-3} 
availability           & \multicolumn{1}{l|}{} & \multicolumn{1}{l|}{http}  \\ \cline{1-1} \cline{3-3} 
botnet                 & \multicolumn{1}{l|}{} & \multicolumn{1}{l|}{https} \\ \cline{1-1} \cline{3-3} 
CDN                    & \multicolumn{1}{l|}{} & \multicolumn{1}{l|}{ftp}   \\ \cline{1-1} \cline{3-3} 
certificate            & \multicolumn{1}{l|}{} & \multicolumn{1}{l|}{ftps}  \\ \cline{1-1} \cline{3-3} 
certificates           & \multicolumn{1}{l|}{} & \multicolumn{1}{l|}{udp}   \\ \cline{1-1} \cline{3-3} 
client                 & \multicolumn{1}{l|}{} & \multicolumn{1}{l|}{lte}   \\ \cline{1-1} \cline{3-3} 
cloud                  & \multicolumn{1}{l|}{} & \multicolumn{1}{l|}{wifi}  \\ \cline{1-1} \cline{3-3} 
communication protocol &                       &                            \\ \cline{1-1}
\end{tabular}%
\end{adjustbox}
\label{tab:net-keywords}
\end{table}

\begin{table}[]
\centering
\caption{The table shows a list of the related CWEs pertaining to network security vulnerabilities. The "CWE ID" is a unique weakness identifier, and "CWE Name" provides more information on the identifier.}
\resizebox{\columnwidth}{!}{
\begin{tabular}{|c|l|}
\hline
\textbf{CWE ID} & \multicolumn{1}{c|}{\textbf{CWE Name}}                                                 \\ \hline
20              & Improper Input Validation                                                              \\ \hline
79              & Improper Neutralization of Input During Web Page Generation ('Cross-site   Scripting') \\ \hline
80              & Improper Neutralization of Script-Related HTML Tags in a Web Page (Basic   XSS)        \\ \hline
83              & Improper Neutralization of Script in Attributes in a Web Page                          \\ \hline
87              & Improper Neutralization of Alternate XSS Syntax                                        \\ \hline
89              & Improper Neutralization of Special Elements used in an SQL Command ('SQL   Injection') \\ \hline
90              & Improper Neutralization of Special Elements used in an LDAP Query ('LDAP   Injection') \\ \hline
91              & XML Injection (aka Blind XPath Injection)                                              \\ \hline
\end{tabular}%
}
\label{tab:net-sec-cwe0}
\end{table}

\noindent\textbf{Attack Graph Partition}.
After obtaining the holistic security posture, another potential need is to get an in-depth security analysis by focusing on the specific layer or scope of the target systems. To achieve this, we propose the attack graph partition component to get subgraphs from the cumulative attack graph.

\sysname follows two approaches to classify the identified vulnerabilities to the layers of interest. In the first approach, \sysname uses a predefined set of high-frequency keywords that typically appear in the corresponding layer. Table~\ref{tab:net-keywords} shows a snippet of the keywords \sysname uses for mapping the vulnerabilities to the network layer. We list the complete list of keywords in Appendix~\ref{appndx:net-sec}. The second approach is based on the MITRE Common Weakness Enumeration (CWE) identifier appointed by the vulnerability database. Table~\ref{tab:net-sec-cwe0} shows a snippet of the list of CWEs \sysname uses for mapping the vulnerabilities to the network layer. We list the complete list in Appendix~\ref{appndx:net-sec}. The reason for using both approaches is that, on the one hand, only keyword matching may result in misclassified vulnerabilities. On the other hand, relying merely on the CWEs is not enough, as many vulnerabilities have not been appointed to CWEs. Additionally, the Dashboard service allows an administrator to change the layer to which a particular vulnerability has been assigned. 
\section{Risk Scoring System}
\label{sec:risk}

Following the construction of attack graphs, a comprehensive assessment of the security posture is conducted through a risk-scoring system (addressing \textbf{C3}). This system receives the attack graphs generated by Graph MS and produces analytics on the security posture.

Within the graphs, each node (i.e., CVE) is linked to exploitability, impact, and risk scores. These scores are derived in accordance with the Common Vulnerability Scoring System (CVSS) standard~\cite{cvss3x} (version 3.1). Recognizing that these scores offer a limited, isolated perspective on vulnerability impact, \sysname enhances the computed scores based on CVSS standards and integrates them for the risk assessment of attack graphs, considering the criticality of affected resources. The evaluation of security posture involves:
\begin{enumerate}
    \item Computation of exploitability, risk, and impact scores for each graph.
    \item Identification of the shortest paths.
    \item Identification of the high-severity paths.
    \item Identification of the key vulnerabilities requiring immediate patching.
    \item Determination of the minimum set of vulnerabilities covering all attack paths.
\end{enumerate}
In what follows, we describe how this risk-scoring system achieves the aforementioned points.

\noindent\textbf{Computing graph exploitability, risk, and impact scores.} The first step is to compute the scores based on the attack graphs. 

\noindent\textit{Edge Exploitability Score (EES).} Let $ees_i$ denote the EES of the edge $i$. It is computed as follows:
\begin{equation} \label{eq:ees}
    ees_i = eScore(source(i)) + c \dot \sum_{x\in in\_edges(source(i))} ees_x
\end{equation}
In the above expression, $source(i)$ is a function that returns the source node of edge $i$ while $eScore()$ returns the exploitability score as a function of the node provided as an argument. The function $in\_edges()$ returns all inbound edges to the node provided as an argument, and $c$ is a predefined multiplication constant ($c=0.1$ in the experiments). Hence, the calculated EES for each edge is associated with the exploitability score of the edge's source node and the scores of all preceding edges in all the paths that include the edge. The computation of $ees_i$ requires all the EES of all edges in the set $in\_edges(source(i))$ to be calculated first.
\\

\noindent\textit{Edge Impact Score (EIS).} Let $eis_i$ denote the EIS of edge $i$. It is computed as follows:
\begin{equation} \label{eq:eis}
    eis_i = iScore(sink(i)) + k \cdot \sum_{x\in out\_edges(sink(i))} eis_x
\end{equation}
In the above expression, $sink(i)$ is a function that returns the sink node of edge $i$ while $iScore()$ returns the impact score as a function of the node provided as an argument. The function $out\_edges()$ returns all the outbound edges from the node provided as an argument, and $k$ ($k=0.01$ in the experiments) is a predefined multiplication constant. Hence, the computed EIS of each edge is associated with the impact score of the edge's sink node and the scores of all the subsequent edges in all the paths that include the edge. In simpler terms, the EIS score of an edge accumulates the impact of all subgraphs starting with that edge.
The computation of $eis_i$ requires all the EIS of all the edges in the set $out\_edges(sink(i))$ to be computed first.

The functions $eScore()$ and $iScore()$ within \sysname are intentionally designed to be user-provided. Formulating a precise function that accurately encapsulates the exploitability and impact scores across diverse deployment scenarios poses a considerable challenge. For instance, the impact of a CVE affecting a device deployed within a critical infrastructure network significantly differs from its impact when the same device is located in a publicly-facing demilitarized zone. To address this complexity, users are granted the flexibility to define their own customized functions. These functions can incorporate the assigned CVE scores along with other parameters of their choosing, allowing for a more tailored and context-aware evaluation of exploitability and impact in varied deployment scenarios. If those functions are not provided, \sysname uses each CVE's assigned CVSS exploitability and impact scores.

\noindent\textit{Edge Risk Score (ERS).} Let $ers_i$ denote the ERS of edge $i$. It is computed as follows:
\begin{equation} \label{eq:ers}
    ers_i = weight_i \cdot (ees_i + eis_i)
\end{equation}
In the above expression, $weight_i$ is the weight of edge $i$, and the values $ees_i$ and $eis_i$ are computed as per the Equations \ref{eq:ees} and \ref{eq:eis}, respectively. Hence, the computed ERS of each edge is associated with the impact and exploitability scores of the edge $i$. The computation of $ers_i$ requires all the EIS and EES of the edge $i$ to be computed first.

Once the computation of the edge scores is completed, we normalize them on a scale of 0 (low) to 10 (high). We apply the normalization for each set of edge scores (i.e., EES, EIS, ERS) for each score as follows. 
\begin{equation}
    normalized\_edge\_score = \dfrac{10 \cdot edge\_score}{max\_set\_score}
\end{equation}
In the above equation, $normalized\_edge\_score$ is the calculated normalized score for a particular edge, $edge\_score$ is the original edge score, and $max\_set\_score$ is the maximum score in the set.

\noindent\textit{Graph Scores.} For each generated graph, 
\sysname computes the graph exploit, impact, and risk scores. Each score is computed as the average of the EES, EIS, and ERS sets.

\noindent\textbf{Identifying the shortest paths with respect to the attacker goals.} We define the shortest path towards the attacker's goals as the path in which the sum of the edge scores present in the path is the highest. Given a graph, we first find the maximum exploitability score of all the nodes in the graph. We, then, define the weight of the edge $i$ as follows:
\begin{equation}
    edge\_weight_i = max\_exploit - eScore(source(i))
\end{equation}
In the above equation, $max\_exploit$ is the maximum exploitability score of all the nodes in the graph. Thus, the higher the exploitability score of the source node of an edge $i$, the lower the weight assigned to the edge. We then add a temporary node to the graph where all the sink nodes (i.e., the attacker goals) have an edge targeting it. This node is called the ``supersink'' node. Finally, we run a weighted shortest path algorithm from the attacker node to the sink node. The algorithm runs in polynomial time and finds the paths with the least weight, thus the highly exploitable paths.

\noindent\textbf{Identifying the high severity attack paths.} To compute all the paths from the attacker node to the attacker's goals, \sysname searches for all the possible paths up to a certain number of edges; this is the cutoff limit for path explorations. For each computed path, \sysname computes the exploitability, impact, and risk score of every path, which is essentially the total sum of EES, EIS, and ERS of the edges in the path. Then \sysname sorts the paths in descending order of risk, exploitability, and impact. \sysname allows the system administrator to change how sorting is performed through the dashboard interface.

\noindent\textbf{Identifying the key vulnerabilities that require immediate patching.} To find those vulnerabilities in the attack graph, we measure the degree of every intermediate node (i.e., any node other than the source or sinks) in the graph. The degree of the node is defined as the number of edges that are incident to the node. 

A node with a high degree essentially means that the node (i.e., CVEs in our attack graphs) is present in several attack paths. Thus, eliminating (i.e., patching) such nodes may render several attacks impractical.

\noindent\textbf{Identifying the minimum set of vulnerabilities that cover all the attack paths.} We apply a minimum set of vertex cover on the constructed attack graph to identify the minimum set of vulnerabilities that cover all the edges in the graph. As vertex cover is an NP-hard problem, we use a local-ratio approximation algorithm to find the minimum set vertex cover~\cite{bar1985local}. Thus, \sysname identifies the minimum set of vulnerabilities that could render the attack paths impractical once mitigated.

\section{Implementation}
\label{sec:implementation}

\begin{figure}[] 
\centerline{\includegraphics[width=1\columnwidth]{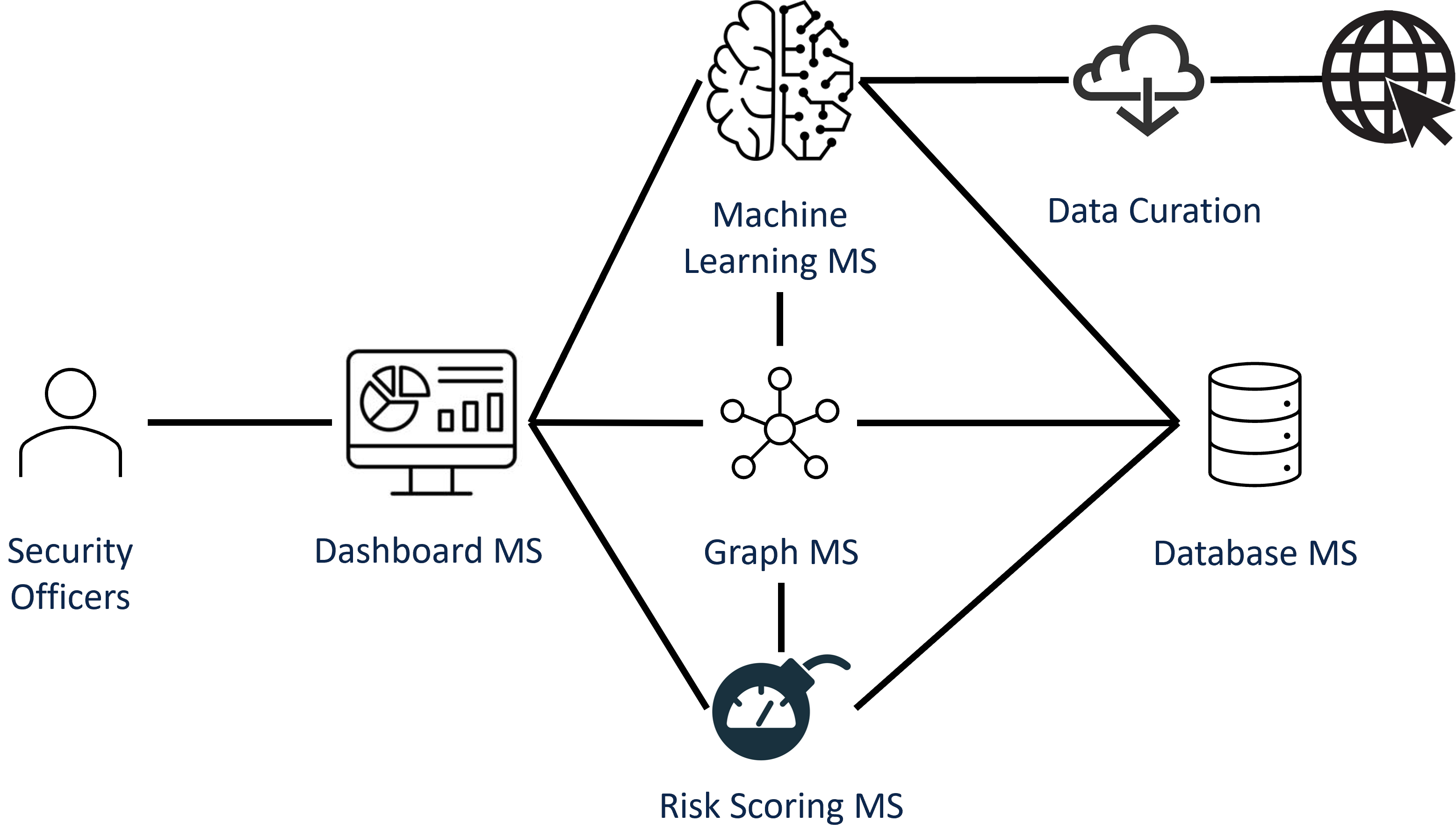}}
\caption{
\sysname is composed of five microservices (MS): the Dashboard MS, Graph MS, Machine Learning MS, Risk Scoring MS, and Database MS.
\label{fig:graph_overview}
}
\end{figure}

We have implemented \sysname of 6,116 lines of Python codes.
\sysname is comprised of five microservices (MS) (see~~\autoref{fig:graph_overview}). Steps 1 and 2 of the pipeline (see~\autoref{fig:pipeline}) are implemented by the Machine Learning MS, step 3 by Graph MS and step 4 by the Risk Scoring MS. Those services are implemented with 987 lines of Python code.
Specifically, we build the named entity recognition and word embedding models based on spaCy~\cite{honnibal2020spacy}, Gensim~\cite{rehurek2011gensim}, and Scikit-learn~\cite{pedregosa2011scikit}.
For the risk scoring system, we use the NetworkX~\cite{hagberg2008exploring} library for graph construction, graph traversal, and risk score computation. \looseness=-1

\begin{figure*}[htbp]
\centerline{\includegraphics[width=1.5\columnwidth]{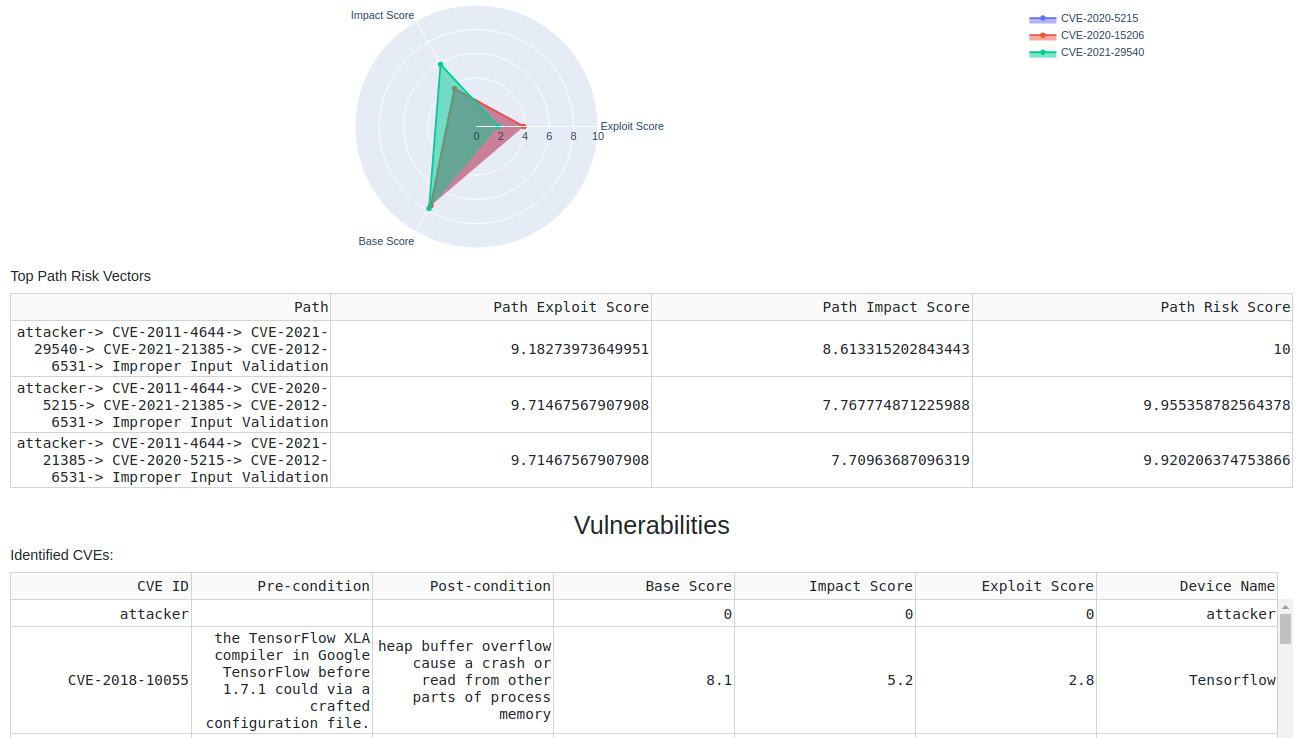}}
\caption{
A small snapshot of \sysname' front-end dashboard interface. 
\label{fig:graphene_dashboard_gen}
}
\end{figure*}

The Dashboard MS orchestrates the entire \sysname pipeline. It has been developed with about 1.5K lines of code using  Flask~\cite{flask} and Dash~\cite{dash} libraries.
The dashboard provides an interface allowing the user (e.g., security officers and network admins) to import details about the network infrastructure, such as the network topology, communicating entities, and device details. Once the posture analysis is completed, the dashboard presents the generated graphs and the results of the risk analysis. For instance, \autoref{fig:graphene_dashboard_gen} shows \sysname' front-end dashboard interface snapshot. The figure showcases a spider graph delineating the top three identified vulnerabilities and subsequently outlines the top three high-severity paths. Additionally, a table excerpt is presented, detailing identified vulnerabilities along with their corresponding extracted pre- and postconditions.

\begin{figure}[htbp]
\centering
\includegraphics[width=0.5\textwidth]{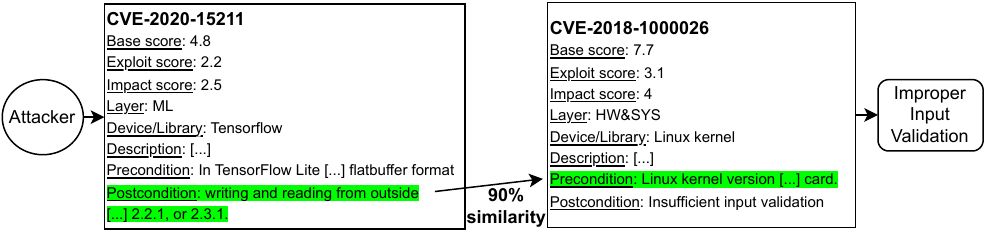}
\caption{An example of a generated attack graph with two intermediate nodes; the first node is an ML layer CVE, and the second is a hardware \& system CVE.}
\label{fig:graph-gen-ex}
\end{figure}

The user can also introduce new nodes in the graph (i.e., new CVEs) and modify or delete existing ones. Such functionality is needed as often online resources are not immediately updated once a certain vulnerability is disclosed. Moreover, the risk, exploitability, and impact scores may not reflect the criticality of the resources in the infrastructure (e.g., overestimated or underestimated scores). Thus, \sysname allows users to change those scores. Once the user makes such changes, the dashboard communicates the request to the back end, where a change-impact analysis occurs. The results are returned to the dashboard, which is immediately updated.

The dashboard also illustrates the generated attack graphs for the given infrastructure for the different layers. For example, Figure~\ref{fig:graph-gen-ex} shows an example of a multi-layer attack graph. The first CVE node is classified under the ML layer, whereas the second CVE under the hardware and systems area.

The dashboard also provides visual representations of the generated attack graphs pertaining to distinct layers within the designated infrastructure. For instance, Figure~\ref{fig:graph-gen-ex} shows a multi-layer attack graph generated by \sysname. In the example in the figure, the initial CVE node is categorized within the Machine Learning (ML) layer, while the subsequent CVE node is categorized within the domain of hardware and systems.

The Databases MS acts as a store and query module. All the constructed graphs and their analysis are stored in the Database MS. This MS uses a high-performance graph database management system (i.e., Neo4j graph database~\cite{miller2013graph}) to expedite querying, storing, and updating operations for large and complex attack graphs.

\section{Evaluation}
\label{sec:eval}

In this section, we first evaluate the effectiveness of our NER-based model for processing CVE disclosures, followed by an evaluation of our word embedding technique used for attack graph construction. We then present a case study reasoning about the output of \sysname.
Our evaluations aim to answer the following research questions:
\begin{itemize}
    \item \textbf{RQ1}: How effective is the named entity recognition model for attack graph nodes?
    \item \textbf{RQ2}: Can the proposed word embedding method semantically match the attack graph nodes?
\end{itemize}
\paragraph{Test Environment} We deploy \sysname on a Google Cloud virtual machine with an Intel Broadwell CPU, 4 GB memory, 110 GB storage, the Ubuntu 20.04 operating system, and an NVIDIA Tesla P100 graphics card. \looseness=-1

\subsection{RQ1: Effectiveness of Named Entity Recognition Model}
\label{sec:ner-effectiveness}
To answer \textbf{RQ1}, we evaluate our named entity recognition (NER)  model by comparing it with baseline models on a large vulnerability NER dataset. 

\paragraph{Dataset} 
Our dataset is adapted from PMA~\cite{guo2022detecting}, which contains descriptions of 52,532 CVEs, 245,573 labeled entities, and 1,828,597 words.
The entities have been annotated and classified into six categories, which are vulnerability type, affected products, root cause, impact, attacker type, and attack vector. It is worth noting that these entities are in line with the vulnerability definitions of the official CVE template~\cite{mitre}.
We split the dataset into training, validation, and test sets based on the 8:1:1 split ratio.


\begin{table}[]
\centering
\caption{Overall Results of Our NER Model and Baselines}
\vspace{-0.1in}
\resizebox{0.454\textwidth}{!}{
\begin{tabular}{lccc}
\toprule
\multicolumn{1}{l}{}                              & \multicolumn{2}{c}{\textbf{Baselines}}                                               & \multicolumn{1}{l}{}                            \\\cmidrule{2-3}
\multicolumn{1}{c}{\multirow{-2}{*}{\textbf{Performance}}} & \textbf{en\_core\_web\_sm} & \textbf{en\_core\_web\_lg} & \multicolumn{1}{l}{\multirow{-2}{*}{\textbf{Our Model}}} \\ 
\midrule
Precision                                         & 95.88                                & 95.2                                 & \textbf{98.75}                                          \\
Recall                                            & 96.07                                & 96.65                                &\textbf{98.55}                                           \\
F1   Score                                        & 95.97                                & 95.92                                & \textbf{98.65 }\\
\bottomrule
\end{tabular}
}
\label{tab:overall-ner}
\end{table}

\begin{table}[]
\centering
\caption{Performance of Our NER Model on Individual Vulnerability Entities}
\begin{tabular}{lrrr}
\toprule
\textbf{Entity}    & \textbf{Precision} & \textbf{Recall} & \textbf{F1 Score} \\
\midrule
Attacker Type     & 97.55              & 98.08           & 97.81             \\
Impact             & 99.59              & 99.59           & 99.59             \\
Attack Vector     & 98.48              & 97.6            & 98.04             \\
Root Cause        & 98.68              & 99.02           & 98.85             \\
Vulnerability Type & 99.03              & 98.35           & 98.69             \\
Affected product  & 98.21              & 98.36           & 98.28            \\
\bottomrule
\end{tabular}
\label{tab:entity-perf}
\end{table}

\paragraph{Baselines and Evaluation Metrics}
Although our NER model encoder is constructed on a transformer encoder, we showcase its superior adaptability by constructing baseline NER models utilizing alternative encoders. 
Specifically, as our NER model is developed atop spaCy~\cite{honnibal2020spacy}, we develop baselines utilizing spaCy's \texttt{en\_core\_web\_sm} and \texttt{en\_core\_web\_lg} encoders~\cite{honnibal2020spacy}.
We train, fine-tune, and evaluate our NER model and baselines on the same training\footnote{The number of training steps is set to 2000, with which our NER model and baselines exhibited loss convergence.}, validation, and test datasets.
We evaluate our NER model and baselines with macro metrics, i.e., precision, recall, and F1 score~\cite{liu2022chinese}.

\paragraph{Evaluation Performance}\autoref{tab:overall-ner} presents the overall performance of the compared models
across all the vulnerability entities.
The results show the superior performance of our NER model compared to baselines.
Moreover, we also evaluate our NER model on individual vulnerability entities.
\autoref{tab:entity-perf} shows the detailed performance, i.e., precision, recall, and F1 score, of our NER model on the individual vulnerability entities.
Among the entities, our NER model performs the best on the ``impact'' entity but the worst on the ``attacker type'' entity.
To understand the performance differences between these two entities, we conducted a manual investigation on our test samples and discovered that vulnerability descriptions often provide more explicit information about the impact rather than the attacker types.
As an illustration, the NVD description of CVE-2017-11341 does not mention the attacker type in its description: ``\textit{There is a heap-based buffer over-read in lexer.hpp of LibSass 3.4.5. A crafted input will lead to a remote denial of service attack.}''.

\subsection{RQ2: Effectiveness of Our Word Embedding Method}
\label{sec:word2vec-effectiveness}

To address \textbf{RQ2}, we perform evaluations on our word embedding methods for semantically matching attack graph nodes.

\paragraph{Word Embedding Dataset and Training} 
Based on the dataset curation framework presented in \autoref{fig:crawler-framework}, we are able to obtain 62,544 pre-processed security documents to train our word embedding models, which are stored in the SQLite database.
The dataset sampling approach (\S\ref{sec:embedding}) results in 6,335,336 word embedding training samples in total.
To model the word semantics in the security context, we choose to train the Continuous Bag-of-Words (CBOW) and Skip-Gram models based on the SentencePiece framework~\cite{kudo2018sentencepiece}.
We train the models to optimal performance by monitoring the loss curve which showed a convergence trend.

\begin{figure}[]
\centering
\includegraphics[width=.8\columnwidth]{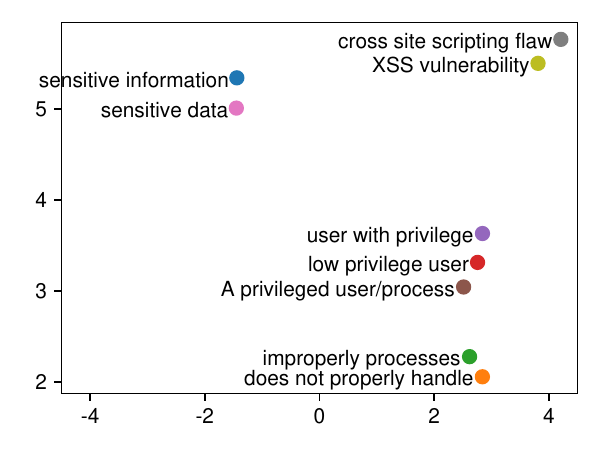}
\vspace{-0.2in}
\caption{t-SNE Evaluation of Generated Embeddings. The distance between phrases shows their semantic similarity.}
\label{fig:tsne-eval}
\end{figure}

\paragraph{Evaluation} 
To evaluate word embedding, the common practice is using the public synonyms datasets, e.g., WS-353 and MTurk-287~\cite{wang2019evaluating}.
However, to the best of our knowledge, there is no such open dataset in the security context.
Additionally, constructing such a dataset with ground truth requires both too much human effort and domain-specific expertise, e.g., linguistic and security knowledge.
Therefore, we opt to visualize word embeddings by using the t-distributed stochastic neighbor embedding (t-SNE)~\cite{van2008visualizing} to evaluate the performance of our word embedding models; 
t-SNE is an algorithm used for data visualization by reducing high-dimensional data to two or three dimensions~\cite{van2008visualizing}. 
It does so by preserving the local structure of the data and creating a low-dimensional map where similar data points are grouped together.
We apply t-SNE to our word embedding models by visualizing the semantic similarity between attack circuit ports, e.g., the input of one attack graph node and the output of another attack graph node.
\autoref{fig:tsne-eval} presents such an example, where the distance of dots representing phrases corresponds to their semantic similarity.
This figure shows that similar semantic phrases are clustered together, such as ``cross site scripting flaw'' and ``XSS vulnerability''. 
The results show that our word embedding models can capture the word semantics well in the security context.

\subsection{Case Study}

\begin{figure}[]
\centering
\includegraphics[width=.7\columnwidth]{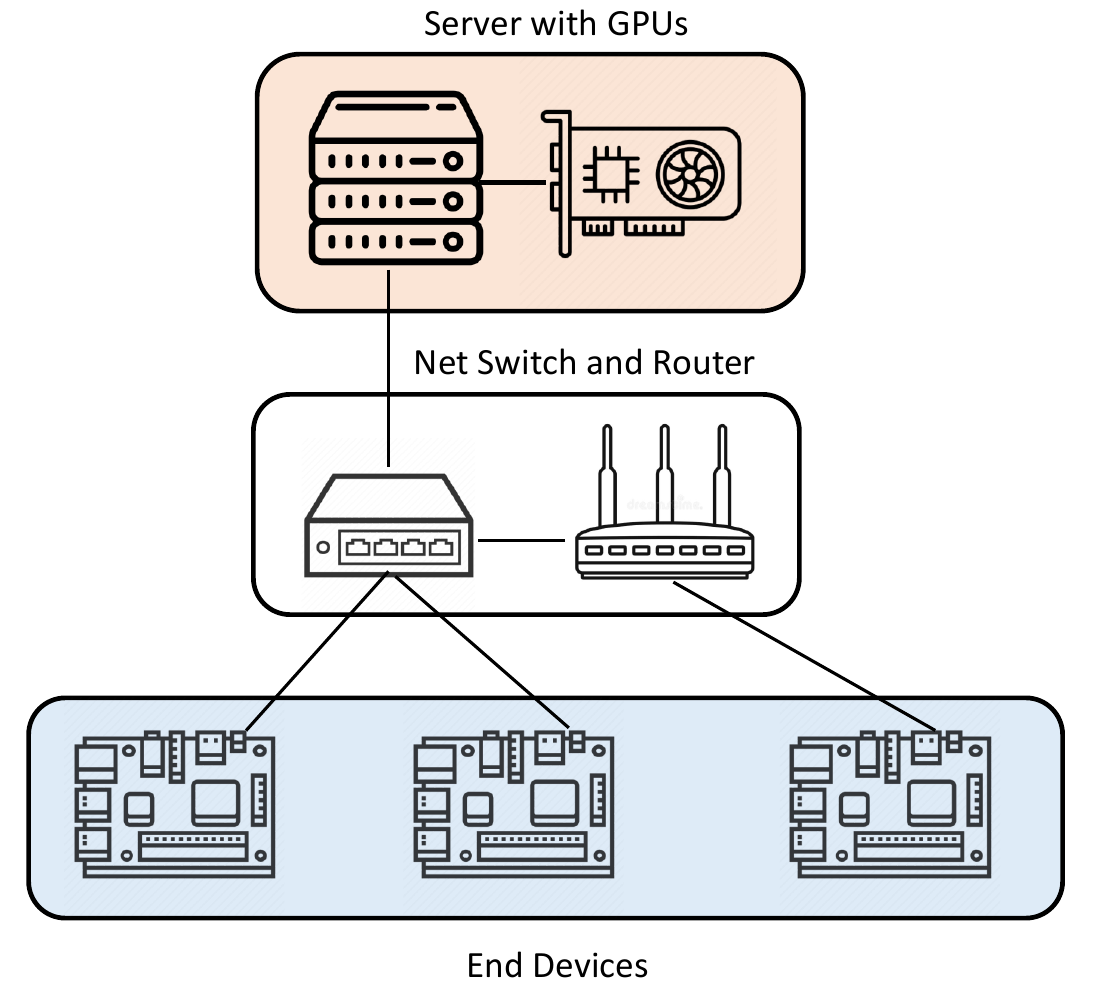}
\vspace{-0.2in}
\caption{Test System for \sysname}
\vspace{-0.2in}
\label{fig:test-system}
\end{figure}

To evaluate \sysname on capturing the holistic security posture of computer systems, we test it on the infrastructure shown in~\autoref{fig:test-system}.
The test system includes three major components: (1) Jetson Nano device as the server with GPUs, (2) TP-link devices as network switch and router, and (3) Raspberry Pi boards as end devices.
Moreover, we also adopt the common communication library, OpenSSL, as the software for securing network traffic. 

To assess the security of the test system, we follow the workflow shown in \autoref{fig:graph_overview}. That is, we first curate all 18K CVEs (till June 2022) from the NVD CVE database. 
We then identify CVEs related to the devices and library used in \autoref{fig:test-system} by matching the product names in CPE entries. After manually confirming the CVEs, we identified 99 CVEs vulnerability descriptions and reports for the test system, including 4 CVEs for Raspberry Pi, 8 CVEs for Jetson Nano, 35 CVEs for TP-link, and 52 CVEs for OpenSSL. 

\begin{figure}[]
	\centering
	\includegraphics[width=.8\columnwidth]{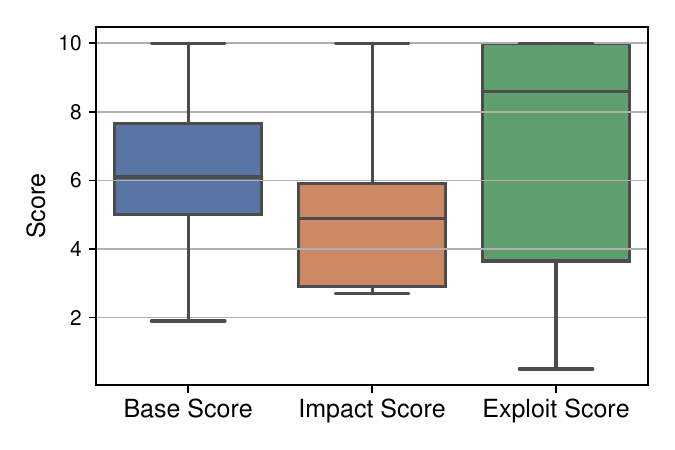}
	\vspace{-0.2in}
	\caption{Distribution of Vulnerability Scores}
        \vspace{-0.2in}
	\label{fig:vuln-score}
\end{figure}

Next, we collect CVE metadata from the NVD database, i.e., CVE descriptions and vulnerability scores. CVE descriptions carry enriched security information. The mean, median, minimal, and maximum number of words in the descriptions are 48.5, 45, 9, and 131, respectively. Moreover, the most frequent bigram and trigram are ``remote attackers'' and ``denial of service'', which appear in the descriptions 38 and 34 times, showing the popular types of attackers and attacks. For risk assessment, \autoref{fig:vuln-score} presents the distribution of CVE base scores, impact scores, and exploit scores.
From this figure, we observe that (1) CVSS base scores show a pretty high risk associated with the CVEs; (2) compared to base scores, the impact scores are in a narrow range; (3) the high impact scores reflect the high likelihood or probability that CVEs will be actively exploited in real-world attacks.

\begin{figure}[]
	\centering
	\includegraphics[width=.8\columnwidth]{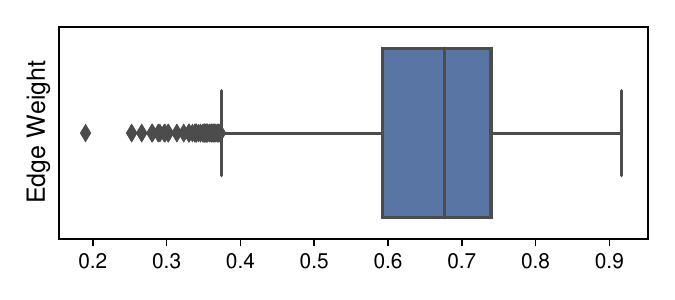}
	\vspace{-0.2in}
	\caption{Distribution of Attack Graph Edge Weights}
	\label{fig:edge-weights}
\end{figure}

\paragraph{Attack Graph Construction} To identify attack graph nodes and build attack graph edges, we run REST APIs of the named entity recognition (NER) and word embedding models. 
The average inference time of the NER and word embedding models are 0.023 and 0.0017 seconds for each input, respectively.
Afterward, we convert the identified entities into the input, output, precondition, and postcondition for each attack graph node. To build the edges, we calculate the Cosine similar score between attack graph nodes based on the embeddings generated by the word embedding models, which serve as the edge weights.
\autoref{fig:edge-weights} presents the distribution of attack graph edge weights. The distribution shows that the majority of attack graph nodes are with high connectivity, i.e., the average edge weight is 0.683. 
In our experiments, we applied a threshold of 0.8 for the similarity score. In practical terms, if CVE \textit{A} has a similarity score lower than 0.8 with CVE \textit{B}, there is no edge connecting them in the resulting attack graphs. Our resulting cumulative attack graph contains 100 graph nodes, including one attacker node, 80 CVE nodes, and 19 target (CWE) nodes.

\paragraph{Risk Scoring.} Table~\ref{tab:risk-results} shows some numerical results as well as performance measures for the scenario illustrated in Figure~\ref{fig:test-system}. We provide the results for the cumulative attack graph as well as the layered attack graphs. As mentioned in Section~\ref{sec:risk}, the exploit, impact, and risk scores are within the [0-10] range. The score computation time (expressed in seconds) includes the time needed to traverse the generated attack graph from the graph service and compute the graph exploitability, risk, and impact scores. The risk analysis time (expressed in seconds) is the time needed to rest the risk scoring system operations, that is, (1) identifying the shortest attack paths and the high-severity paths, (2) identifying key vulnerabilities, and (3) the minimum set of vulnerabilities that cover all the attack paths.

In the cumulative attack graph, \sysname has identified CVE-2020-5215, CVE-2020-15206, and CVE-2021-29540 as the top three critical vulnerabilities associated with the installation of two vulnerable Tensorflow versions (2.4.2 and 1.15.2, respectively) on servers equipped with GPUs, as depicted in the spider graph in~\autoref{fig:graphene_dashboard_gen}. These vulnerabilities are notably part of the computed vertex cover, indicating their presence in the set of vulnerabilities encompassing all attack paths. Additionally, \sysname highlighted the top three attack paths based on their elevated risk, exploitability, and impact, as shown in the ``top path risk vectors'' section of~\autoref{fig:graphene_dashboard_gen}. It is worth noting that in every generated attack graph, at least one node corresponds to the computed vertex cover, identifying the minimum set of vulnerabilities that must be addressed. Addressing these vulnerabilities through patching may render all attack graphs impractical. Ultimately, \sysname has identified the five shortest attack paths within the cumulative attack graph, each having one CVE node with an exploitability score of 10, denoting the highest possible score. These CVE nodes are specifically CVE-2010-3173 (Mozilla Firefox on end devices), CVE-2011-0392 (Cisco TelePresence Recording Server on the server), CVE-2012-6531 (Zend Framework on end devices), CVE-2012-0884 (OpenSSL in all devices and the server), and CVE-2015-0763 (Cisco Unified MeetingPlace on end devices). Notably, analogous observations hold true for the network, system \& hardware, machine learning, and cryptography layers.

\begin{table}[]
\centering
\caption{Risk scoring results for the scenario described in Figure~\ref{fig:test-system}.}
\vspace{-0.1in}
\resizebox{\columnwidth}{!}{%
\begin{tabular}{|l|c|c|c|c|c|}
\hline
\textbf{Layer} &
  \textbf{Commulative} &
  { \textbf{Network}} &
  { \textbf{System \& HW}} &
  { \textbf{ML}} &
  { \textbf{Crypto}} \\ \hline
Exploit score (/10)    & 2.89  & 3.57 & 3.068  & 5.74 & 2.94  \\ \hline
Impact score (/10)     & 3.07  & 3.18 & 3.8322 & 5.82 & 4.277 \\ \hline
Risk score (/10)       & 2.31  & 2.72 & 3.9583 & 5.41 & 2.61  \\ \hline
Total nodes            & 100   & 27   & 30     & 30   & 22    \\ \hline
Number of attack paths & 27297 & 137  & 53     & 6289 & 20    \\ \hline
Shortest attack paths  & 5     & 3    & 4      & 3    & 2     \\ \hline
Vertex cover size      & 50    & 19   & 15     & 21   & 3     \\ \hline
{ \begin{tabular}[c]{@{}l@{}}Score computation time\\ (seconds)\end{tabular}} &
  0.0198 &
  0.0018 &
  0.0014 &
  0.0051 &
  0.00098 \\ \hline
{ \begin{tabular}[c]{@{}l@{}}Risk analysis time \\ (seconds)\end{tabular}} &
  1.1874 &
  0.0053 &
  0.0021 &
  0.2538 &
  0.00097 \\ \hline
\end{tabular}%
}
\label{tab:risk-results}
\end{table}
\section{Related Work}
\label{sec:related}


\noindent \textbf{Attack graph construction.} For attack graph construction, prior approaches focused primarily on addressing the key challenge of extracting attack information~\cite{swiler1998graph, aksu2018automated,inokuchi2019design,weerawardhana2015automated,payne2019secure,weerawardhana2014automated,jones2015towards,li2019self,simran2020deep}. These approaches can be categorized as rule-based and learning-based methods.
For rule-based methods~\cite{aksu2018automated,inokuchi2019design,weerawardhana2015automated,payne2019secure}, all rules are manually defined, limiting their scalability. 
Learning-based methods are built upon traditional machine learning algorithms~\cite{weerawardhana2014automated,jones2015towards}, such as support vector machine, or deep learning models~\cite{li2019self,simran2020deep}.
However, such learning-based methods are usually trained on fuzzy or domain-specific features but fail to learn the general vulnerability semantics in the security context.
For example, these previous approaches have employed pre-trained word vectors that were trained on general English datasets. However, these methods tend to be less accurate in the security domain due to the presence of specific terminology and linguistic semantics that differ from those found in general English repositories.

\noindent\textbf{Named Entity Recognition and Word Embedding for Security Applications}.
Natural language processing (NLP) has been widely used for security applications, e.g., binary analysis~\cite{jin2023binary,ding2019asm2vec} and security patch detection~\cite{wu2022enhancing}.
The security applications of NLP typically involve the utilization of named entity recognition (NER) and word embedding techniques.
The application of NER to the descriptions and analysis of vulnerabilities is not new, and previous works have analyzed vulnerabilities from different aspects with domain-specific NER~\cite{weerawardhana2014automated, binyamini2021framework,jin2022understanding}.
For example, Weerawardhana et al.~\cite{weerawardhana2014automated} propose identifying security concepts in vulnerability information focusing on home computer security.
Binyamini et al.~\cite{binyamini2021framework} design a framework to model the attack techniques from security vulnerability descriptions, where the attack surfaces are the main focuses (e.g., software, hardware, operating systems, and network ports).
However, there are several key limitations in those approaches, including (1) their scope is limited, e.g., home computer~\cite{weerawardhana2014automated} and attack surfaces~\cite{binyamini2021framework}, and (2) they use limited machine learning models, such as LSTM networks~\cite{binyamini2021framework} and convolutional neural networks~\cite{guo2022detecting}.

Word embedding generates distributed representations of natural language words for efficient computations~\cite{mikolov2013distributed,mikolov2013efficient}.
Similar to its applications in natural language processing, existing research has adopted it to security applications~\cite{shen2019attack2vec}.
For example, Shen et al.~\cite{shen2019attack2vec} model the cyberattack steps by temporal word embedding.
Srivastava et al.~\cite{srivastava2023study} proposes enhancing security NLP models by word embeddings.
Corizzo et al.~\cite{corizzo2020feature} learn the raw data representations of network traffic for intrusion detection with word embeddings.
Unlike existing works, we focus on contextualizing the vulnerability-specific word embeddings by training models on massive vulnerability reports, i.e., CVE descriptions.

\noindent \textbf{Risk Scoring.} Li et al.~\cite{li2006cluster} have proposed a cost/benefit analysis of the attack graphs. Administrators have to manually provide cost and benefit values for every node in the attack graph. A cost may relate to the adversary in terms of the resources that they need to possess, whereas a benefit is for the organization. Lu et al.~\cite{ lu2009ranking} use the graph neural networks (GNNs) model to rank attack graphs. The authors use Google's PageRank web search algorithm~\cite{bianchini2005inside} to infer each step's importance in the attack graph.
 \sysname, however, extends the CVSS scoring system, which therefore aligns with the assessment and expertise of the CVE assigner. 
Idika et al.~\cite{ idika2010extending} have proposed attack graph-based security metrics to analyze attack graphs. However, their analysis is focused on the shortest paths and the number of paths an attacker can take to violate a security policy and not on the severity of the attack paths.

Other relevant approaches to security metrics are the probabilistic ones~\cite{abraham2014cyber,liu2005network,wang2008attack}. Abraham et al.~\cite{ abraham2014cyber} use stochastic modeling techniques based on Markov chains to analyze the network's current and future security state. 
Liu et al.~\cite{liu2005network} propose an approach based on Bayesian networks to model potential attack paths. However, for such an approach to work users have to provide the edge exploitability probability for each CVE. 
However, this approach may provide inaccurate analyses due to an over- or underestimated probability assignment. In addition, both of those approaches do not consider the attack's impact on the network infrastructure in terms of resource criticality.

\section{Conclusion and Future Work}
\label{sec:conclusion}

The unceasing growth of cybercrime and network complexity require innovative cybersecurity solutions. \sysname, the security analyzer proposed in this paper, is a promising step toward meeting this demand. Through the use of machine learning, natural language processing, and systematic vulnerability analysis, \sysname offers a comprehensive security solution for general computing infrastructures in that it is capable of extracting semantic meanings from disclosed vulnerabilities, performing layered classification, and creating exhaustive directed attack graphs for holistic security analytics.

The distinctive multi-layered approach to security posture analysis provided by \sysname holds the potential for a more nuanced understanding of a network's security landscape, allowing vulnerabilities of different natures and their interrelations to be effectively addressed. Furthermore, the proposed scoring methods provide a more tailored and relevant security assessment for the infrastructure under analysis, thereby enhancing the practical value of the analytics.

While the application of \sysname shows promise, there is significant scope for future work.
First, improving the machine learning models used in \sysname will be crucial. As these models evolve, the accuracy and effectiveness of semantic extraction of vulnerabilities can be enhanced, leading to a more reliable and detailed vulnerability analysis.
Second, enhancing the granularity of the layer classification process is an interesting research direction. While \sysname classifies vulnerabilities into layers, refining this classification to include sub-layers or specialized categories might provide more targeted insights for security measures.
Third, developing advanced scoring methods can be a pivotal area of future exploration. As \sysname traverses the generated graph, devising innovative and more granular metrics to gauge the severity of vulnerabilities could make risk assessment more precise.

Furthermore, \sysname could be extended to incorporate automated patching and response capabilities to enhance its practical utility. This would allow for immediate remediation of identified vulnerabilities, enhancing the network's overall security. Integrating \sysname with real-time monitoring systems could also provide dynamic, up-to-date security posture analytics, thereby enabling faster response times to potential threats. 
Last, real-world testing of \sysname across various industries, like healthcare, energy, or finance, would provide invaluable feedback for fine-tuning the system. These tests could also uncover specific industry-related vulnerabilities, which could be incorporated into the model for a more sector-specific approach.

Overall, \sysname offers significant strides in cybersecurity. Still, its evolution is a continuous journey that will shape the contours of a safer and more secure digital landscape.




\bibliographystyle{IEEEtranS}
\bibliography{sample-base}

\appendix

\section{Keywords and CWE Lists for layer matching}
\label{appndx:net-sec}


In this section, we list a subset of keywords used by \sysname to classify the vulnerabilities to one of our proposed layers; the Network Security layer. Table~\ref{tab:net-sec-appndx} lists the network-related keywords, whereas Table~\ref{tab:proto-appndx} lists a subset of widely adopted communication protocols in network infrastructures.

\begin{table*}[]
\centering
\caption{The ``Keywords'' column corresponds to high-frequency keywords found in network vulnerability listings. }
\begin{adjustbox}{width=0.4\textwidth,center}
\begin{tabular}{|l|l}
\hline
\multicolumn{1}{|c|}{\textbf{Keywords}} & \multicolumn{1}{c|}{\textbf{Keywords}} \\ \hline
access control                          & \multicolumn{1}{l|}{MITM}              \\ \hline
authentication                          & \multicolumn{1}{l|}{nat}               \\ \hline
authenticity                            & \multicolumn{1}{l|}{NAT}               \\ \hline
authorization                           & \multicolumn{1}{l|}{network}           \\ \hline
availability                            & \multicolumn{1}{l|}{network interface} \\ \hline
botnet                                  & \multicolumn{1}{l|}{network packets}   \\ \hline
CDN                                     & \multicolumn{1}{l|}{packets}           \\ \hline
certificate                             & \multicolumn{1}{l|}{port}              \\ \hline
certificates                            & \multicolumn{1}{l|}{ports}             \\ \hline
client                                  & \multicolumn{1}{l|}{privacy}           \\ \hline
cloud                                   & \multicolumn{1}{l|}{protocol}          \\ \hline
communication protocol                  & \multicolumn{1}{l|}{remote attacker}   \\ \hline
confidentiality                         & \multicolumn{1}{l|}{remote attackers}  \\ \hline
Cross-site request forgery              & \multicolumn{1}{l|}{repudiation}       \\ \hline
Cross-site scripting                    & \multicolumn{1}{l|}{request}           \\ \hline
CSRF                                    & \multicolumn{1}{l|}{response}          \\ \hline
DDoS                                    & \multicolumn{1}{l|}{router}            \\ \hline
denial of service                       & \multicolumn{1}{l|}{sase}              \\ \hline
DoS                                     & \multicolumn{1}{l|}{SDN}               \\ \hline
downgrade                               & \multicolumn{1}{l|}{server}            \\ \hline
edge network                            & \multicolumn{1}{l|}{side-channel}      \\ \hline
edge nodes                              & \multicolumn{1}{l|}{spoof}             \\ \hline
endpoints                               & \multicolumn{1}{l|}{spoofing}          \\ \hline
firewall                                & \multicolumn{1}{l|}{SQL}               \\ \hline
flood                                   & \multicolumn{1}{l|}{switch}            \\ \hline
flooding                                & \multicolumn{1}{l|}{tamper}            \\ \hline
html                                    & \multicolumn{1}{l|}{tampering}         \\ \hline
ICN                                     & \multicolumn{1}{l|}{trust}             \\ \hline
injection                               & \multicolumn{1}{l|}{verification}      \\ \hline
input sanitization                      & \multicolumn{1}{l|}{VPN}               \\ \hline
input validation                        & \multicolumn{1}{l|}{wireless}          \\ \hline
integrity                               & \multicolumn{1}{l|}{XSS}               \\ \hline
IoT                                     & \multicolumn{1}{l|}{zero-trust}        \\ \hline
LAN                                     & \multicolumn{1}{l|}{ZTA}               \\ \hline
man-in-the-middle                       & \multicolumn{1}{l|}{link}              \\ \hline
message                                 & \multicolumn{1}{l|}{network}           \\ \hline
mirai                                   &                                        \\ \cline{1-1}
\end{tabular}%
\end{adjustbox}
\label{tab:net-sec-appndx}
\end{table*}

\begin{table*}[]
\centering
\caption{The protocols are used for communication between entities.}
\begin{adjustbox}{width=0.2\textwidth,center}
\begin{tabular}{|l|l|}
\hline
\textbf{Protocol} & \textbf{Protocol} \\ \hline
tls               & mqtt              \\ \hline
ssl               & coap              \\ \hline
tcp               & amqp              \\ \hline
ip                & lora              \\ \hline
http              & zigbee            \\ \hline
https             & WEP               \\ \hline
ftp               & WPA               \\ \hline
ftps              & icmp              \\ \hline
udp               & tor               \\ \hline
lte               & i2p               \\ \hline
wifi              & TELNET            \\ \hline
bluetooth         & DHCP              \\ \hline
ARP               & DNS               \\ \hline
\end{tabular}%
\end{adjustbox}
\label{tab:proto-appndx}
\end{table*}

\begin{table*}[]
\centering
\caption{The table shows a list of the related CWEs pertaining to network security vulnerabilities. The ``CWE ID'' is a unique weakness identifier, and ``CWE Name'' provides more information on the identifier.}
\resizebox{\textwidth}{!}{%
\begin{tabular}{|c|l|}
\hline
\textbf{CWE ID} & \multicolumn{1}{c|}{\textbf{CWE Name}}                                                 \\ \hline
20              & Improper Input Validation                                                              \\ \hline
79              & Improper Neutralization of Input During Web Page Generation ('Cross-site   Scripting') \\ \hline
80              & Improper Neutralization of Script-Related HTML Tags in a Web Page (Basic   XSS)        \\ \hline
83              & Improper Neutralization of Script in Attributes in a Web Page                          \\ \hline
87              & Improper Neutralization of Alternate XSS Syntax                                        \\ \hline
89              & Improper Neutralization of Special Elements used in an SQL Command ('SQL   Injection') \\ \hline
90              & Improper Neutralization of Special Elements used in an LDAP Query ('LDAP   Injection') \\ \hline
91              & XML Injection (aka Blind XPath Injection)                                              \\ \hline
93              & Improper Neutralization of CRLF Sequences ('CRLF Injection')                           \\ \hline
97              & Improper Neutralization of Server-Side Includes (SSI) Within a Web Page                \\ \hline
98  & Improper Control of Filename for Include/Require Statement in PHP Program   ('PHP Remote File Inclusion') \\ \hline
113 & Improper Neutralization of CRLF Sequences in HTTP Headers ('HTTP   Request/Response Splitting')           \\ \hline
183             & Permissive List of Allowed Inputs                                                      \\ \hline
184             & Incomplete List of Disallowed Inputs                                                   \\ \hline
200             & Exposure of Sensitive Information to an Unauthorized Actor                             \\ \hline
209             & Generation of Error Message Containing Sensitive Information                           \\ \hline
213             & Exposure of Sensitive Information Due to Incompatible Policies                         \\ \hline
269             & Improper Privilege Management                                                          \\ \hline
282             & Improper Ownership Management                                                          \\ \hline
284             & Improper Access Control                                                                \\ \hline
285             & Improper Authorization                                                                 \\ \hline
286             & Incorrect User Management                                                              \\ \hline
287             & Improper Authentication                                                                \\ \hline
287             & Improper Authentication                                                                \\ \hline
288             & Authentication Bypass Using an Alternate Path or Channel                               \\ \hline
289             & Authentication Bypass by Alternate Name                                                \\ \hline
290             & Authentication Bypass by Spoofing                                                      \\ \hline
294             & Authentication Bypass by Capture-replay                                                \\ \hline
295             & Improper Certificate Validation                                                        \\ \hline
296             & Improper Following of a Certificate's Chain of Trust                                   \\ \hline
297             & Improper Validation of Certificate with Host Mismatch                                  \\ \hline
298             & Improper Validation of Certificate Expiration                                          \\ \hline
299             & Improper Check for Certificate Revocation                                              \\ \hline
301             & Reflection Attack in an Authentication Protocol                                        \\ \hline
302             & Authentication Bypass by Assumed-Immutable Data                                        \\ \hline
303             & Incorrect Implementation of Authentication Algorithm                                   \\ \hline
304             & Missing Critical Step in Authentication                                                \\ \hline
305             & Authentication Bypass by Primary Weakness                                              \\ \hline
306             & Missing Authentication for Critical Function                                           \\ \hline
307             & Improper Restriction of Excessive Authentication Attempts                              \\ \hline
308             & Use of Single-factor Authentication                                                    \\ \hline
322             & Key Exchange without Entity Authentication                                             \\ \hline
345             & Insufficient Verification of Data Authenticity                                         \\ \hline
346             & Origin Validation Error                                                                \\ \hline
352             & Cross-Site Request Forgery (CSRF)                                                      \\ \hline
359             & Exposure of Private Personal Information to an Unauthorized Actor                      \\ \hline
385             & Covert Timing Channel                                                                  \\ \hline
417             & Communication Channel Errors                                                           \\ \hline
419             & Unprotected Primary Channel                                                            \\ \hline
420             & Unprotected Alternate Channel                                                          \\ \hline
\end{tabular}%
}
\label{tab:net-sec-cwe}
\end{table*}

\begin{table*}[]
\centering
\caption{The table shows a list of the related CWEs pertaining to network security vulnerabilities. The ``CWE ID'' is a unique weakness identifier, and ``CWE Name'' provides more information on the identifier.}
\resizebox{\textwidth}{!}{%
\begin{tabular}{|c|l|}
\hline
\textbf{CWE ID} & \multicolumn{1}{c|}{\textbf{CWE Name}}                                                 \\ \hline
425             & Direct Request ('Forced Browsing')                                                     \\ \hline
441             & Unintended Proxy or Intermediary ('Confused Deputy')                                   \\ \hline
497             & Exposure of Sensitive System Information to an Unauthorized Control   Sphere           \\ \hline
515             & Covert Storage Channel                                                                 \\ \hline
522             & Insufficiently Protected Credentials                                                   \\ \hline
564             & SQL Injection: Hibernate                                                               \\ \hline
566             & Authorization Bypass Through User-Controlled SQL Primary Key                           \\ \hline
593             & Authentication Bypass: OpenSSL CTX Object Modified after SSL Objects are   Created     \\ \hline
599             & Missing Validation of OpenSSL Certificate                                              \\ \hline
601             & URL Redirection to Untrusted Site ('Open Redirect')                                    \\ \hline
603             & Use of Client-Side Authentication                                                      \\ \hline
611             & Improper Restriction of XML External Entity Reference                                  \\ \hline
613             & Insufficient Session Expiration                                                        \\ \hline
614             & Sensitive Cookie in HTTPS Session Without 'Secure' Attribute                           \\ \hline
638             & Not Using Complete Mediation                                                           \\ \hline
639             & Authorization Bypass Through User-Controlled Key                                       \\ \hline
643             & Improper Neutralization of Data within XPath Expressions ('XPath   Injection')         \\ \hline
644             & Improper Neutralization of HTTP Headers for Scripting Syntax                           \\ \hline
645             & Overly Restrictive Account Lockout Mechanism                                           \\ \hline
652             & Improper Neutralization of Data within XQuery Expressions ('XQuery   Injection')       \\ \hline
706             & Use of Incorrectly-Resolved Name or Reference                                          \\ \hline
776 & Improper Restriction of Recursive Entity References in DTDs ('XML Entity   Expansion')                    \\ \hline
836             & Use of Password Hash Instead of Password for Authentication                            \\ \hline
862             & Missing Authorization                                                                  \\ \hline
863             & Incorrect Authorization                                                                \\ \hline
918             & Server-Side Request Forgery (SSRF)                                                     \\ \hline
923             & Improper Restriction of Communication Channel to Intended Endpoints                    \\ \hline
924 & Improper Enforcement of Message Integrity During Transmission in a   Communication Channel                \\ \hline
939             & Improper Authorization in Handler for Custom URL Scheme                                \\ \hline
940             & Improper Verification of Source of a Communication Channel                             \\ \hline
941             & Incorrectly Specified Destination in a Communication Channel                           \\ \hline
942             & Permissive Cross-domain Policy with Untrusted Domains                                  \\ \hline
1004            & Sensitive Cookie Without 'HttpOnly' Flag                                               \\ \hline
1211            & Authentication Errors                                                                  \\ \hline
1214            & Data Integrity Issues                                                                  \\ \hline
1220            & Insufficient Granularity of Access Control                                             \\ \hline
1263            & Improper Physical Access Control                                                       \\ \hline
1270            & Generation of Incorrect Security Tokens                                                \\ \hline
1275            & Sensitive Cookie with Improper SameSite Attribute                                      \\ \hline
1311            & Improper Translation of Security Attributes by Fabric Bridge                           \\ \hline
1327            & Binding to an Unrestricted IP Address                                                  \\ \hline
1331            & Improper Isolation of Shared Resources in Network On Chip (NoC)                        \\ \hline
1385            & Missing Origin Validation in WebSockets                                                \\ \hline
\end{tabular}%
}
\label{tab:net-sec-cwe2}
\end{table*}

\end{document}